\documentclass[aps,accepted=2017-07-12]{quantumarticle}

\usepackage[SquareTraceBrackets]{quantum}
\usepackage{amsmath}
\usepackage{eufrak}
\usepackage{tikz}
\usepackage[colorlinks=true]{hyperref}

\newcommand{\bmu}{\boldsymbol{\mu}}
\newcommand{\bxi}{\boldsymbol{\xi}}
\newcommand{\ba}{\boldsymbol{a}}
\newcommand{\bL}{\boldsymbol{\mathcal{L}}}
\newcommand{\bX}{\boldsymbol{X}}
\newcommand{\btheta}{\boldsymbol{\theta}}

\newcommand{\change}{\color{black}}

\begin{document}

\title{Optimal Quantum Metrology\\ of Distant Black Bodies}

\author{Mark E. Pearce}\email{mrkprc1@gmail.com}
\affiliation{Department of Physics \& Astronomy, University of Sheffield, Sheffield S3 7RH, United Kingdom}

\author{Earl T. Campbell}\email{earltcampbell@gmail.com}
\affiliation{Department of Physics \& Astronomy, University of Sheffield, Sheffield S3 7RH, United Kingdom}

\author{Pieter Kok}\email{p.kok@sheffield.ac.uk}
\affiliation{Department of Physics \& Astronomy, University of Sheffield, Sheffield S3 7RH, United Kingdom}

\begin{abstract}
\noindent Measurements of an object's temperature are important in many disciplines, from astronomy to engineering, as are estimates of an object's spatial configuration.  We present the quantum optimal estimator for the temperature of a distant body based on the black body radiation received in the far-field.  We also show how to perform {\change separable} quantum optimal estimates of the spatial configuration of a distant object, i.e. imaging.  In doing so we necessarily deal with multi-parameter quantum estimation of incompatible observables, a problem that is poorly understood.  We compare our optimal observables to the two mode analogue of lensed imaging and find that the latter is far from optimal, even when compared to measurements which are separable.  To prove the optimality of the estimators we show that they minimise the cost function weighted by the quantum Fisher information---this is equivalent to maximising the average fidelity between the actual state and the estimated one.

\end{abstract}

\maketitle

\section{Introduction}
\noindent 
The quantum Fisher information (QFI) is a prevalent figure of merit in the field of quantum parameter estimation \cite{Escher2012,Macieszczak2014,Kok2015,Jarzyna2015,Knott2016}.  The quantum analogue of the Cram\'er-Rao bound provides a lower bound to the covariance matrix of the parameters to be estimated, although the attainability of this bound is not guaranteed as for classical statistics \cite{Helstrom1976}.  As in the classical case \cite{Kay1993}, the quantum Fisher information for states that follow a Gaussian distribution takes on a closed form that depends only on the first and second moments and their derivatives \cite{Monras2013,Gao2014,Safranek2015}.  The fact that thermal states are so commonly found in nature and exhibit Gaussian statistics makes them the ideal testbed for Gaussian quantum estimation problems \cite{Marian2016,Monras2011}.  

The Cram\'er-Rao bound gives a lower bound on the covariance matrix of unbiased parameter estimates by the inverse of the Fisher information matrix \cite{Kay1993}.  Physically, this describes the relationship between the information obtained about a parameter via a measurement and the uncertainty in an estimate of the parameter from the measurement data.  Similarly, the quantum Cram\'er-Rao bound gives a lower bound on estimation uncertainty via the inverse of the QFI \cite{Helstrom1967}.  Since the QFI is a property of the state alone and does not depend on a particular measurement scheme, the precision in parameter estimates is determined by the uncertainty in the state only \cite{Braunstein1994}, and is therefore fundamental in nature and cannot be reduced by improving measurement apparatus.

In classical metrology, the extension to multiple parameters does not entail a significant alteration to the basic theory.  However, in the quantum theory, the possibility of observables being incompatible leads to additional complications \cite{Szczykulska2015}.  The task of finding optimal observables under such circumstances is far from trivial and may involve collective measurements over many independent copies of the system, which is experimentally challenging.  In this paper we consider only separable measurements (each system measured independently) and attempt to find the optimal observables among this class of observables.

Thermometry is an important method for interrogating the physical world. The temperature of astronomical objects reveals important clues about their nature, e.g. the cosmic microwave background \cite{Fixsen2009} and estimates of effective stellar temperatures \cite{Alonso1999}; in both engineering and living systems, the temperature of components provides an essential diagnostic tool. The temperature of an object can be measured in multiple ways \cite{Moldover2016}, for example via estimating the heat flow in direct-contact measurements, or by remotely measuring the radiation field emitted by the object. Typically, the thermal radiation field emitted is a black body spectrum.  In this paper we consider the latter method, and present optimal quantum estimators for the temperature of black body emitters.

Imaging provides us with important information about the spatial configuration of an object.  Typically, imaging techniques are not studied for their optimality.  Even in high resolution imaging, the focus tends to be towards increasing the resolution beyond the diffraction limit and rarely are optimal estimators considered.  The main obstacle in applying the powerful techniques of metrology (both classical and quantum) to imaging is in identifying the parameters that constitute an image.  Optimal metrology for an object of known parameterisation can be performed \cite{Tsang2015, Tsang2016, Pearce2015a, Lupo2016} but for an object with an unknown spatial configuration it is not obvious what parameters we need to estimate.  We show that the spatial configuration of a thermally radiating source is determined by spatial correlations in the far-field.  In this paper we show how to optimally estimate these parameters, which completely determine the state, therefore mapping the problem of imaging to state estimation.  

{\change The density matrix of blackbody radiation is constructed from tensor products of many independent thermal ${\bf k}$-modes \cite{MandelandWolf1995}:
\begin{align}\label{eq:rho_def}
\rho = \underset{{\bf k}}{\otimes} \rho_{\bf k} \, ,
\end{align}
where the tensor product runs over all ${\bf k}$ of the form ${\bf k} = \Delta {\bf k} (x,y,z)$ where $x,y,z$ are non-negative integers.  Note we coarse grain ${\bf k}$-space rather than working with a continuum of modes.  Physically, we motivate this coarse graining by appealing to the finite size of any realisable detector.}  

The spatial properties of the source are usually unknown, and without knowledge of them we typically cannot make accurate temperature measurements.  A simple example helps to clarify the problem: consider the estimation of a star's temperature.  If we simply measure the photon count at a single frequency in the far-field, we cannot distinguish between hotter sources and those with a larger angular size since an increase in either parameter will lead to a larger photon count.  Therefore, if we wish to measure the temperature of the source, we must either know the angular size of the source or attempt to estimate it {\change by measuring more }than one frequency mode. 

If we are interested in the temperature of the source alone, then the radiating area---or more precisely the solid angle of the source---is a nuisance parameter.  Regardless of whether we are interested in the solid angle of the source, it is necessary to estimate it in order to provide optimal estimates of the temperature.  The converse is also {\change generally} true.  However, as we will show, it is not possible to make measurements of {\change arbitrary} spatial properties of the source by measuring spectral modes alone.  In fact, it is necessary to consider spatially separated modes, where measurements of the correlations provide information about the spatial configuration of the source.      

A blackbody emits at all frequencies with an intensity determined by the Planck distribution.  This distribution can be considered as an infinite number of independent spectral modes with a spectral width inversely proportional to the time over which the state is observed \cite{MandelandWolf1995}.  It is therefore quite straightforward to determine the statistics of these independent modes and calculate the quantum Fisher information to determine the optimal measurements for simultaneous temperature and solid angle estimates.  Considering multiple spatial modes is somewhat more cumbersome as the spatial modes are not generally independent.  Here, by appealing to the fundamental quantum mechanical description of blackbody radiation, we demonstrate that it is the correlations between modes that convey the information about the spatial configuration of the source to the far-field. 

\begin{figure}[t]
\centering
\begin{tikzpicture}[]
 \node at (0,0){\includegraphics[width=1\columnwidth]{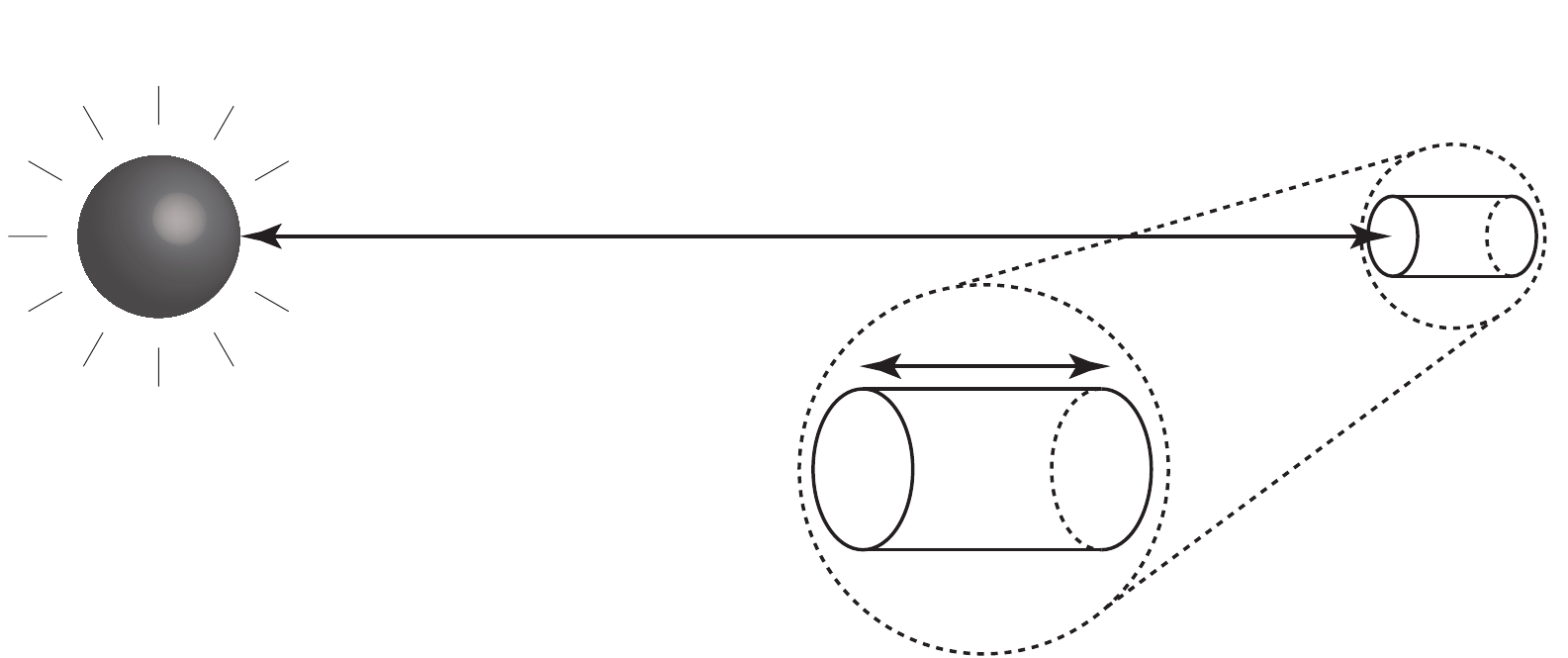}};
 \node at (-3.2,-0.6){\small Black body};
 \node at (-0.5,0.8){\small $R$};
 \node at (1.05,-0.75){\small $\rho$};
 \node at (1.05,0){\small $c \tau$};
 \node at (0.45,-0.75){\small $A_{\rho}$};

\end{tikzpicture}
\caption{The definition of our state $\rho$.  We assume that the transverse area over which the state is measured, $A_{\rho}$, is much smaller than the coherence area of the radiation.  We also assume that the time over which the state is measured, $\tau$, defines the spectral width $\Delta \nu = \tau^{-1}$.}\label{fig:rho_def}
\end{figure}
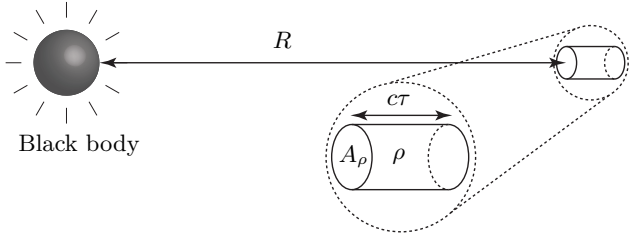

It has been stated before that the optimal measurement for temperature estimates is photon number counting, \cite{Helstrom1968,Nair2015}.  However, these results rely upon the assumption that we can measure a single thermal mode in the far-field, an assumption that we claim cannot be satisfied without knowledge of the exact spatial properties of the source.  This is due to the transverse coherence area of the far-field radiation being determined by the spatial properties of the source \cite{MandelandWolf1995}, therefore to guarantee we are measuring exactly one mode, our detectors must be designed with these properties in mind.  In this paper, the state $\rho$ we consider is defined in Fig.~\ref{fig:rho_def}.  {\change The state describes the radiation occupying a small volume $A_{\rho} \times c \tau$ in the far-field of a black body source, where $A_{\rho}$ is the transverse area of the state and $\tau$ is the observation time.}

An important assumption is that the transverse area of the state $A_{\rho}$ is much smaller than the coherence area of the source.  The effect is that we observe a fraction of each spectral mode, therefore ensuring that the transverse coherence is approximately equal to one across the entire area $A_{\rho}$.  This allows us to coarse grain in the transverse direction, ignoring any effects due to the finite transverse coherence area of the radiation, within a single spatial mode.  

In Sec.~\ref{sec:GSF} we review the Gaussian state formalism, giving the definition of these states and establishing the notation we will use throughout this paper.  Sec.~\ref{sec:QFI} introduces the quantum Fisher information and gives an outline of the general principles of quantum metrology.  In Sec.~\ref{sec:Temp_est} we determine the density matrix for a blackbody state observed in the far-field and determine the optimal measurements for the parameters that govern these states.  Unsurprisingly these turn out to be photon counting.  In Sec.~\ref{sec:TSM} we consider the effect of adding a secondary spatial mode, transversely separated from the first.  We find it important to consider the coherence between the spatial modes, and in fact this becomes the essential parameter to estimate if we wish to glean spatial information about the source.  We consider how to optimally obtain this spatial information and discover that the optimal observables do not commute.  Therefore we are forced to find a measurement that can be considered optimal whilst being restricted by the inherent uncertainty of incompatible observables.  A precisely optimal separable POVM is found that depends upon the values of the parameters.  We also find a POVM which is independent of the parameter values and yet is close to optimal, which is of great practical importance. 

\section{Gaussian State Formalism}\label{sec:GSF}
\noindent Gaussian states are defined to be states with a Gaussian characteristic function, which for an $n$-mode state defined by the bosonic mode operators $\ba = \smash{(\hat{a}_1,\hat{a}^{\dagger}_1,\dots,\hat{a}_n,\hat{a}^{\dagger}_n)^{\rm T}}$, has the form \cite{Gao2014}
\begin{align}
\chi(\bxi) = \tr{\rho {\rm e}^{-\ba^{\rm T} \Omega \bxi}} = \exp\left( \frac{1}{2}(\Omega \bxi)^{\rm T}  \Sigma  \Omega \bxi +  \bxi \Omega \bmu \right) \, ,
\end{align}
where $\bxi = (\xi_1, \xi_1^*,\dots , \xi_n, \xi_n^* )^{\rm T}$, $\Omega = \oplus_{i=1}^n i \sigma_y$, {\change $\sigma_y$ is the Pauli $y$ matrix,} and $\bmu$ and $\Sigma$ are the first and second moments defined by
\begin{align}
\mu^{\alpha} & = \braket{a^{\alpha}} \, ,\\
\Sigma^{\alpha \beta} & = \frac{1}{2}\left(\braket{\lambda^{\alpha} \lambda^{\beta}} + \braket{\lambda^{\beta} \lambda^{\alpha}}\right) \, ,
\end{align}
and $\lambda^{\alpha} = a^{\alpha} - \mu^{\alpha}$.  Often the quantities we wish to calculate can be given in closed form as a function of these moments.  Due to the central limit theorem, Gaussian states are frequently encountered in systems where there are a large number of randomly fluctuating influences \cite{Holevo2011,Hudson1973}.

\section{Quantum Fisher Information}\label{sec:QFI}

\noindent In classical parameter estimation, the Cram\'er-Rao bound states that the covariance matrix of unbiased estimates of the parameters, $\btheta = (\theta_1, \dots, \theta_d)^{\rm T}$, is bounded by the inverse of the Fisher information matrix \cite{Kay1993}.  That is,
\begin{align}\label{eq:CRB}
\boldsymbol{\Sigma}_{\btheta} =  \braket{ (\hat{\btheta} - \braket{\hat{\btheta}}) (\hat{\btheta} - \braket{\hat{\btheta}})^{\rm T}} \geq {\bf I}_C^{-1} \, ,
\end{align}
where the hat over the $\btheta$ signifies that we are concerned with estimates of the parameters, and not the parameters themselves (and should not be confused with a quantum mechanical operator), and ${\bf I}_C$ is the classical Fisher information matrix:
\begin{align}\label{eq:CFI}
{\bf I}_C = \sum_{\rm x} \frac{\left(\nabla_{\btheta} p({\rm x}| \btheta) \right) \left( \nabla_{\btheta} p({\rm x}| \btheta) \right)^{\rm T}}{p({\rm x}| \btheta)} \, ,
\end{align}
where $\nabla_{\btheta} = (\frac{\partial}{\partial \theta_1}, \dots , \frac{\partial}{\partial \theta_d})^{\rm T}$.  The sum in Eq.~\eqref{eq:CFI} is performed over all possible outcomes, ${\rm x}$, of the conditional probability distribution $p({\rm x}| \btheta)$.

The matrix inequality of Eq.~\eqref{eq:CRB} should be understood as implying that the matrix $\boldsymbol{\Sigma}_{\btheta} - {\bf I}_C^{-1}$ is positive semi-definite.  It follows that the inequality
\begin{align}
\tr{ G \, \boldsymbol{\Sigma}_{\btheta}} \geq \tr{ G \,{\bf I}_C^{-1}} \, ,
\end{align}
also holds for any positive definite weight matrix $G$.  The weight matrix $G$ allows us to assign a relative importance to different parameters. 

The bound in Eq.~\eqref{eq:CRB} expresses a limitation of classical data processing.  It states that estimates of a set of parameters, $\btheta$, are constrained in their variance by a quantity that is entirely determined by the probability distribution from which we sample data, $\{ {\rm x} \}$.  In quantum metrology, we consider how the covariance matrix of our estimates is constrained if the state, $\rho(\btheta)$, is measured by some self-adjoint  operator, $\smash{\hat{X} = \sum_{\rm x} {\rm x} \, \hat{\Pi}_{\rm x}}$, giving rise to the probability distribution $p({\rm x}| \btheta) = \tr{\rho(\btheta) \hat{\Pi}_{\rm x}}$.  We assume that we are capable of measuring any self-adjoint operator, which allows us to reduce the variance by changing $p({\rm x}| \btheta)$.  Finding the observable that minimises the variance is therefore the objective of quantum metrology.

In analogy with classical metrology, we can define a quantum Fisher information matrix, ${\bf I}_Q$, which leads to a quantum version of the Cram\'er-Rao bound.  For a single parameter, the quantum Fisher information is given by the expectation of the square of $\mathcal{L}$, the so-called symmetric logarithmic derivative (SLD):
\begin{align}
 {\bf I}_Q = \tr{\rho \mathcal{L}^2}\, .
\end{align}
The SLD is implicitly defined by
\begin{align}
 \partial_{\theta} \rho = \frac{\rho \mathcal{L} +  \mathcal{L} \rho }{2}  \, .
\end{align}
In the multi-parameter case, this generalises to a quantum Fisher information matrix with elements \cite{Paris2009}
\begin{align}\label{eq:QFI}
{\bf I}_Q =  {\rm Re}\left( \tr{\rho  \bL  \bL^{\rm T}} \right)\, ,
\end{align}
where we have defined $\bL = (\mathcal{L}_1, \dots, \mathcal{L}_d)^{\rm T}$, which is the vector of SLD operators, and $\mathcal{L}_i$ is defined in analogy to the single parameter case $\frac{1}{2} (\rho \mathcal{L}_i  + \mathcal{L}_i \rho ) = \partial_{\theta_i} \rho$.  For brevity, we write $\partial_i = \partial_{\theta_i}$ throughout the remainder of the paper.  

In \cite{Gao2014} Gao and Lee derived a closed form for the QFI of a Gaussian state in terms of the first and second moments, $\bmu$ and $\Sigma$.  For blackbody states $\mu = 0$ and the expression found in \cite{Gao2014} reduces to
\begin{align}
{\bf I}_Q & = \frac{1}{2} \mathfrak{M}^{-1}_{\alpha \beta, \gamma \kappa} \Big( \nabla_{\btheta} \Sigma^{\gamma \kappa} \Big) \Big(  \nabla_{\btheta}\Sigma^{\alpha \beta} \Big)^{\rm T}\, ,
\end{align}
where the matrix $\mathfrak{M} = \Sigma \otimes \Sigma +\frac{1}{4} \Omega \otimes \Omega$, and the summation convention is used for greek indices.  Gao and Lee also provide an expression for the SLD, which is given in terms of $\mathfrak{M}$:
\begin{align}
\mathcal{L}_i = \frac{1}{2} \mathfrak{M}^{-1}_{ \gamma \kappa,\alpha \beta} \left( \partial_i \Sigma^{\alpha \beta} \right)(a^{\gamma} a^{\kappa} - \Sigma^{\gamma \kappa}) \, ,
\end{align}
which we will make use of later to calculate the SLDs for blackbody states.  

Eq.~\eqref{eq:QFI} gives the so-called SLD form of the quantum Fisher information.  A quantum version of Eq.~\eqref{eq:CRB} can be determined, which, for a single parameter, is known to be attainable, at least in the asymptotic sense \cite{Barndorff2000}, {\change by which we mean that it is possible to find an estimator where the variance of estimates tends towards the inverse of the QFI as the number of independent measurements tends to infinity}.  

To attain the bound requires finding a measurement for which the classical Fisher information is equal to the quantum Fisher information.  For a single parameter, the optimal measurement is related to the SLD.  However, in general the SLD will depend on the exact value of the parameter, which is assumed unknown before any measurements are made.  The conventional method to circumvent this issue is to perform a sub-optimal measurement first to obtain an initial estimate of the parameter, then use this estimate to measure an approximately optimal operator, preferably adaptively changing the measurement as more information is obtained about the parameters true value \cite{Barndorff2000,Bagan2006,Gill2000,Giovannetti2011}.  

For multiple parameters, the bound is not always attainable because the optimal observables may not commute and therefore may not be simultaneously measurable \cite{Helstrom1976}.  This introduces a new problem, independent to the problem of the optimal measurements depending on the true values of the parameters.  Even if the SLDs do not commute, the quantum Cram\'er-Rao bound is asymptotically attainable if the following condition is satisfied \cite{Ragy2016}:
\begin{align}\label{eq:expcomm}
\tr{\rho [\mathcal{L}_i, \mathcal{L}_j]} = 0 .
\end{align}
However, to achieve the quantum Cram\'er-Rao bound in this context assumes the use of a collective measurement on multiple independent copies of the state.  Although this is interesting, the implementation of such a measurement may well prove difficult.  In Sec.~\ref{sec:TSM} we encounter a problem with non-commuting SLDs and attempt to find the optimal separable measurement for our system. 

\section{Temperature Estimates of Far-Field Blackbody Sources}\label{sec:Temp_est}

\noindent In this section we show how the quantum Fisher information can be calculated for temperature estimates of far-field blackbody radiation.  We also find the optimal estimators for both temperature and solid angle and show that the quantum Cram\'er-Rao bound is attainable since the SLDs commute.

Blackbody radiation exhibits a Gaussian characteristic function and can therefore be analysed with the preceding formalism by Gao and Lee.  In addition, blackbody states also have zero first moment, i.e., $\mu^{\alpha} = 0$ for all $\alpha$.  We assume that we observe the radiation from a blackbody state in the far-field, paraxial regime.  Considering only a single spectral mode centred on the frequency $\nu$, we find that the state has covariance matrix 
\begin{align}
\Sigma = \left( \braket{n_{\nu}} +\frac{1}{2} \right) \sigma_x\, ,
\end{align}
{\change where $\sigma_x$ is the Pauli $x$ operator.  The covariance matrix depends only on $\braket{n_{\nu}}$ and we} can therefore, estimate at most a single parameter.  

From Planck's law, we can calculate the average number of photons for the arrangement given in Fig.~\ref{fig:rho_def},
\begin{align}
\braket{n_{\nu}} = \frac{ \nu^2 A_S A_{\rho}}{2 \pi c^2 R^2 } \frac{1}{{\rm e}^{\beta h \nu} - 1} = \nu^2 \kappa \braket{n_{\rm th}} \, ,
\end{align}
where $A_S$ is the radiating area of the source, $\braket{n_{\rm th}} = [\exp(\beta h \nu) -1 ]^{-1}$ is the average number of photons per thermal mode, and $\kappa$ contains all of the parameters regarding the arrangement of the source and its relation to the state $\rho$.  We can also interpret $\nu^2 \kappa$ as the number of modes of frequency $\nu$ that are captured by the state $\rho$.  As mentioned above, we assume throughout that $\nu^2 \kappa \ll 1$.  

Given that we do not know the value of $\kappa$ beforehand, the only way we can ensure that this condition is satisfied is by using appropriate values for the parameters that we can control, $A_{\rho}$ and $\nu$, the frequency we choose to observe.  For example, using the area of a single CCD pixel as $A_{\rho}$, $\sim10^{-11}{\rm m}^2$, we find that this condition is equivalent to {\change $\cot \vartheta \gg \alpha \nu$, where $\vartheta$ is the angular size of the source and $\alpha=10^{-14} {\rm s}^{-1}$}.  This condition should easily be satisfied in the far-field regime for frequencies up to X-rays.

Identifying $\theta_1 = T$ and $\theta_2 = \kappa$ we find that the QFI matrix for the spectral mode centred on $\nu$ is given by
\begin{align}\label{eq:QFISM}
{\bf I}_Q^{(\nu)} & =  \frac{1}{\braket{n_{\nu}} + \braket{n_{\nu}}^2}  \begin{pmatrix}
(\partial_1 \braket{n_{\nu}})^2 & \partial_1 \braket{n_{\nu}} \partial_2 \braket{n_{\nu}} \\
\partial_2 \braket{n_{\nu}} \partial_1 \braket{n_{\nu}} & (\partial_2 \braket{n_{\nu}})^2 
\end{pmatrix} \nonumber \\
& = \frac{1}{\braket{n_{\nu}} + \braket{n_{\nu}}^2} \nabla_{\btheta} \braket{n_{\nu}} (\nabla_{\btheta} \braket{n_{\nu}})^{{\rm T}},
\end{align}
which is a rank 1 projector and therefore rank deficient. It is thus a singular matrix that does not have a well-defined inverse. Consequently, it is not possible to estimate both parameters with finite precision.  We conclude that conforming to our expectation, we cannot find optimal unbiased estimators for both parameters from a single mode.  However, if for some reason we do know one of the parameters precisely, we can make an estimate of the other.  The SLD in that case is given by
\begin{align}\label{eq:SLD1M}
\mathcal{L}_i = \frac{ \partial_i \braket{n_{\nu}} }{\braket{n_{\nu}} + \braket{n_{\nu}}^2} \hat{a}^{\dagger} \hat{a} - \frac{ \braket{n_{\nu}} \partial_i \braket{n_{\nu}} }{\braket{n_{\nu}} + \braket{n_{\nu}}^2} \id \, ,
\end{align}
and the quantum Cram\'er-Rao bound for the single parameter $\theta_i$ is $\braket{(\Delta \theta_i)^2} \geq [\braket{n_{\nu}} +\braket{n_{\nu}}^2]/(\partial_i \braket{n_{\nu}})^2$.  This result was obtained in Ref. \cite{Helstrom1968} and shows that the optimal measurement for temperature estimation is photon counting measurements.  This follows from the form of the SLD in Eq.~\eqref{eq:SLD1M} and the observation that the optimal estimator for $\theta_i$ is given by \cite{Paris2009}
\begin{align}\label{eq:Xop}
X = \frac{\mathcal{L}_i}{{\rm I}_Q} \, .
\end{align}
Measurements of $X$ are therefore achieved by measuring in the basis of the SLD $\mathcal{L}_i$.  Since this is the number basis we can perform photon number counting and postprocess the outcomes to find the value of $\theta_i$. 

When we are ignorant of both parameters, which will generally be the case, we are required to make use of at least two spectral modes.  The covariance matrix then becomes
\begin{align}\label{eq:2Spec_Modes}
\Sigma = \overset{M}{\underset{ i=1}{\oplus} }\left( \braket{n_{\nu_i}} + \frac{1}{2} \right)	 \sigma_x \, ,
\end{align}
where $M$ is the total number of spectral modes we consider.  Since each spectral mode is independent, the QFI matrix is given by
\begin{align}
{\bf I}_Q = \sum_{i=1}^M {\bf I}_Q^{(\nu_i)}  \, ,
\end{align}
where $\smash{{\bf I}_Q^{(\nu_i)}}$ are given by matrices of the form Eq.~\eqref{eq:QFISM}.  Despite the matrices ${\bf I}_Q^{(\nu_i)} $ being singular, the matrix ${\bf I}_Q $ is non-singular as long as $M \geq 2$.  The SLD for $\theta_i$ is given by
\begin{align}\label{eq:SLD1}
\mathcal{L}_i & =  \sum_{j=1}^M \frac{(\partial_i \braket{n_{\nu_j}}) (\braket{n_{\nu_j}} - \hat{n}_{\nu_j})}{ \braket{n_{\nu_j}} + \braket{n_{\nu_j}}^2 }  \, .
\end{align}
From Eq.~\eqref{eq:SLD1} we see that $[\mathcal{L}_1,\mathcal{L}_2]=0$ and therefore it is possible to measure in the eigenbasis of both SLDs simultaneously.  Therefore, photon counting in each of the $M$ independent spectral modes provides an optimal estimate of both parameters simultaneously.  {\change In practice this is typically how stellar temperatures are measured, although often more sophisticated models are used to allow for deviations away from exact black body behaviour.}  
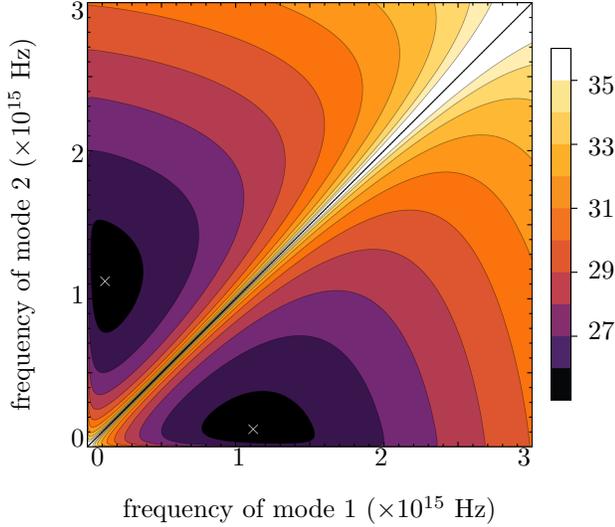
\begin{figure}[t]
\begin{tikzpicture}[]
  \node at(-0.38,0) {\includegraphics[width=0.7\columnwidth]{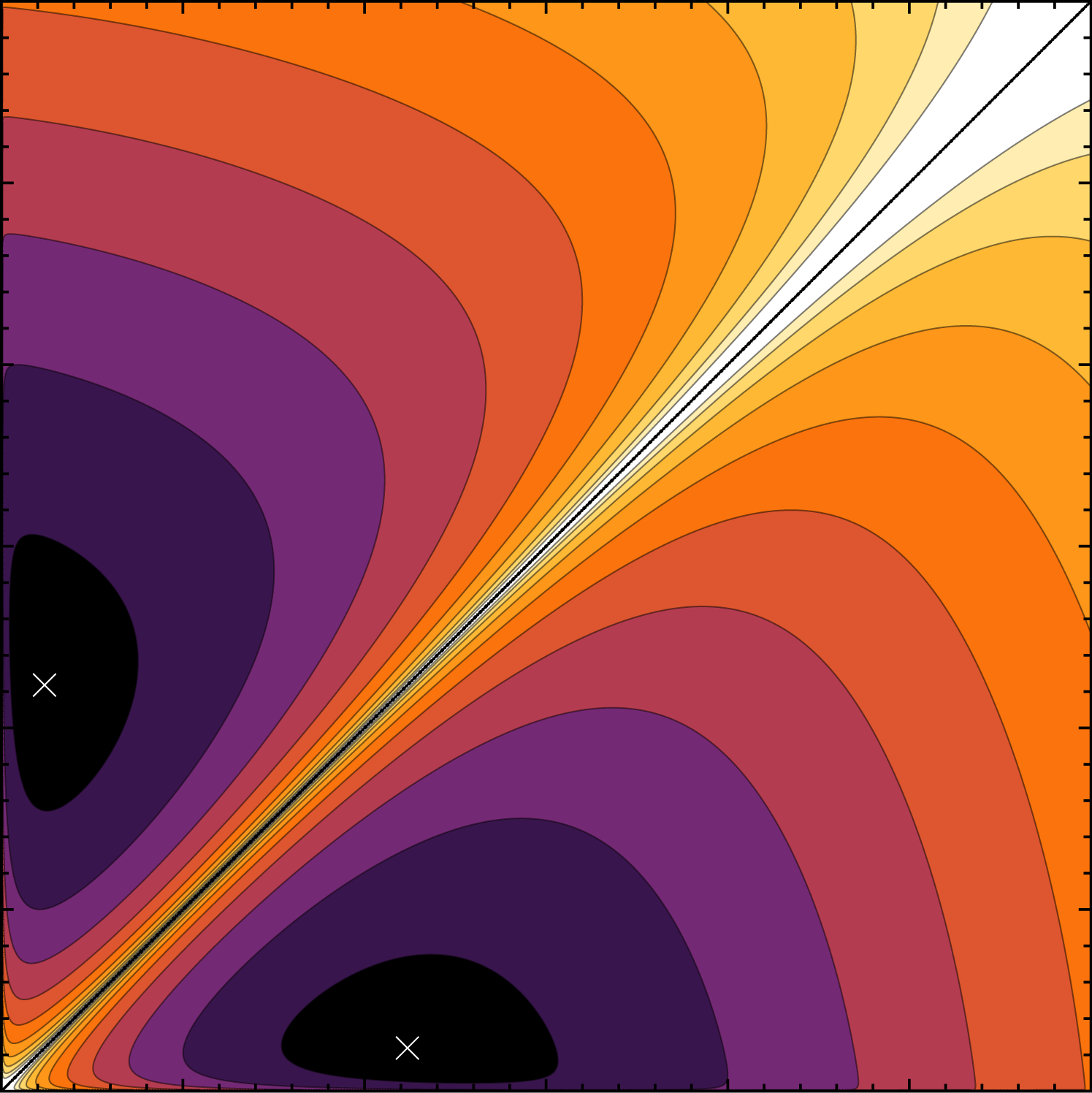}};
  \node at(3,0) {\includegraphics[width=0.04\columnwidth]{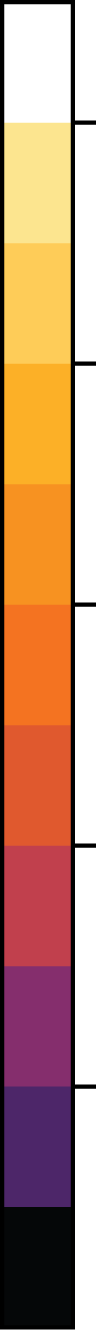}};
  
  \node at (-3.2,-3.1){0};
  \node at (-1.32,-3.1){1};
  \node at (0.57,-3.1){2};
  \node at (2.47,-3.1){3};
  \node at (-0.38,-3.8){frequency of mode 1  ($\times 10^{15}$ Hz)};
  
  \node[anchor=east] at (-3.25,-2.85){0};
  \node[anchor=east] at (-3.25,-0.94){1};
  \node[anchor=east] at (-3.25,0.94){2};
  \node[anchor=east] at (-3.25,2.84){3};
  \node [rotate=90] at (-4.2,0){frequency of mode 2  ($\times 10^{15}$ Hz)};
  
  \node[anchor=west] at (3.2,-1.42) {27};
  \node[anchor=west] at (3.2,-0.61) {29};
  \node[anchor=west] at (3.2,0.21) {31};
  \node[anchor=west] at (3.2,1.03) {33};
  \node[anchor=west] at (3.2,1.85) {35};
  
\end{tikzpicture}
\caption{The natural logarithm of the variance of $T$ (in units of Kelvin) as a function of the frequencies $\nu_1$ and $\nu_2$.  The exact location of the optimal frequencies is dependent on the value of $T$.  In this plot we have used $T = 10000$ K and $\kappa = 10^{-32} \, s^{2}$, which ensures that $N \ll 1$ over the range of frequencies plotted.  It is not surprising that the values on the bar are so large when we consider that the variance quoted is for a single round of measurement and we are using only two spectral modes.  Also, at 10000 K the average number of photons per measurement is always less than 0.0003.  The white crosses show the location of the minimum.}\label{fig:varT}
\end{figure}

We stress that photon counting must be performed on at least two spectral modes in order to obtain estimates for both parameters.  The variance for the parameter $\theta_i$ is
\begin{align}
& \braket{(\Delta \theta_i)^2 }   \geq \nonumber \\
& \frac{ \sum_{l=1}^M \left[ {\bf C}^{(\nu_l)} \right]_{ii} }{\sum_{l,k=1}^M \left( \left[ {\bf I}^{(\nu_l)}_Q \right]_{11} \left[ {\bf I}^{(\nu_k)}_Q \right]_{22} - \left[ {\bf I}^{(\nu_l)}_Q \right]_{12} \left[ {\bf I}^{(\nu_k)}_Q \right]_{21} \right)}\, ,
\end{align}
where ${\bf C}^{(\nu_l)}$ is the cofactor matrix of $ {\bf I}^{(\nu_l)}_Q$, and ${\bf I}^{(\nu_l)}_Q  $ is  
\begin{align}
{\bf I}^{(\nu_l)}_Q = \frac{( \nabla_{\btheta} \braket{n_{\nu_l}}) ( \nabla_{\btheta}\braket{n_{\nu_l}})^{\rm T}}{\braket{n_{\nu_l}} + \braket{n_{\nu_l}}^2} \, .
\end{align}
Assuming that we are interested in estimating $T$ and treating $\kappa$ as a nuisance parameter, we plot in Fig.~\ref{fig:varT} the variance in estimates of $T$ for two spectral modes as a function of the frequency of each mode.  In Fig.~\ref{fig:varT} we see that there is a well defined optimum for the frequencies $\nu_1$ and $\nu_2$, the exact value of which we would expect to depend upon the precise values of the parameters to be estimated.  

Numerically we find that the optimal values of $\nu_1$ and $\nu_2$ are independent of $\kappa$ and linearly dependent on $T$, approximately $\nu_1^{({\rm min})} = (1.188 \times 10^{10} {\rm \, Hz \, K^{-1}}) T$, $\nu_2^{({\rm min})} = (1.118 \times 10^{11} {\rm \, Hz \, K^{-1}} )T$.  This is reminiscent of Wien's displacement law, which gives the peak frequency of blackbody radiation as $\nu^{({\rm max})} = (5.88\times10^{10}  {\rm \, Hz \, K^{-1}}) T$.  In the absence of an initial estimate of $T$, the best we can do is to perform photon counting for two frequencies and adjust these correspondingly as information is obtained about the temperature.  If we happen to know of an a priori distribution for the temperature (for example, when estimating the temperature of stars an a priori distribution can be obtained using observational data about the relative frequency of stellar temperatures), we can choose the initial frequencies based upon minimisation of the averaged variance, that is
\begin{align}
\min_{\nu_1, \, \nu_2} \int  \, {\rm d}T \, p(T) \, [{\bf I}_Q^{-1}]_{11} \, ,
\end{align}
where $p(T)$ is our prior distribution over $T$.

Generally we may want to estimate both parameters.  A higher precision can be achieved in our estimates by increasing the number of spectral modes observed.  Assuming that we are only capable of photon counting in $M$ spectral modes simultaneously, an optimal measurement can be found given these limited resources. Defining $\boldsymbol{\nu} = (\nu_1 , \dots , \nu_M)^{\rm T}$, an optimal measurement is given by
\begin{align}
\min_{\boldsymbol{\nu}} \tr{G \, {\bf I}^{-1}_Q (\boldsymbol{\nu})} \, .
\end{align}
The solution of this minimisation will depend on the actual values of the parameters $T$, and $\kappa$, and the chosen weight matrix $G$.  Therefore, without prior knowledge about the parameters, an optimal measurement for a given weight matrix is one that minimises
\begin{align}
\min_{\boldsymbol{\nu}} \int {\rm d} \kappa \, {\rm d} T \, p(\kappa, T) \tr{G \, {\bf I}^{-1}_Q (\boldsymbol{\nu})} \, ,
\end{align}
where $p(\kappa, T)$ is a prior distribution over the parameters, which will depend strongly upon the exact application.

We are now naturally led to the question: what measurement can we perform to determine the spatial configuration of the source?  {\change Most single parameter problems can be treated by estimating $\kappa$, which can then be used to determine this single parameter}.  For example if the source is simply planar circular, then the estimation of the parameter $\kappa \propto A_S/R^2 = 2 \pi r^2/R^2 = 2 \pi \vartheta^2$ is equivalent to estimating the angular radius $\vartheta$.  In general however, since the QFI depends only upon the average number of photons in each frequency mode, which in turn depends only on the source temperature and $\kappa$, we cannot estimate general spatial properties of the source, no matter how many spectral modes we measure.  
\begin{figure}[t]
\centering
\begin{tikzpicture}[]
\pgftext{\includegraphics[width=0.8\columnwidth]{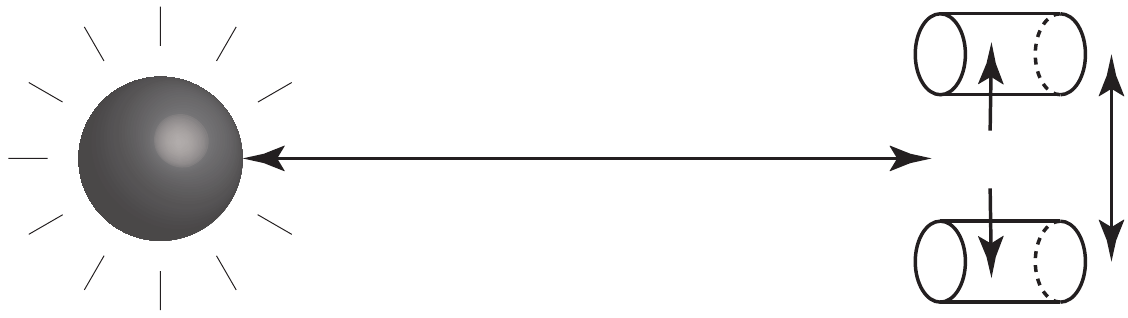}} at (-2,0);
 \node at (-2.3,-1.2){ Black body};
 \node at (0,0.3){ $R$};
 \node at (2.45,0){ $\rho$};
 \node at (3.4,0){ $x$};
\end{tikzpicture}
\caption{The addition of a secondary spatial mode, the modes are transversely separated by a distance $x = |{\bf x}_1 - {\bf x}_2|$.  The state $\rho$ is now a composite system consisting of two spatially separated modes.  We again assume that the transverse area of each spatial mode is less than the coherence area of the radiation, which means that each mode captures a fraction of a spectral mode.  The van Cittert-Zernike theorem allows us to calculate the far-field complex degree of coherence between the two modes, $\gamma({\bf x}_1 , {\bf x}_2) = \gamma_{12}$.}\label{fig:state2}
\end{figure}
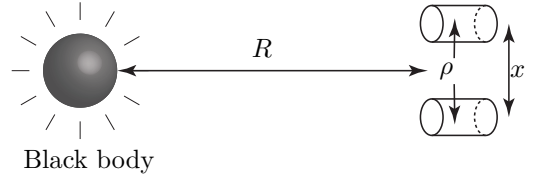

Since the spatial properties are all contained within the parameter $\kappa$, any attempt to estimate more than a single spatial parameter results in a singular QFI matrix.  In order to allow the estimation of arbitrary spatial properties, we need to consider the effect of adding additional spatial modes, which are spatially separated from the first mode, see Fig.~\ref{fig:state2}. 

\section{{\change Estimation of Spatial Parameters}}\label{sec:TSM}

\noindent {\change In the previous section we showed how to optimally estimate the temperature of black body sources.}  In this section we show that the addition of spatial modes allows for the determination of the spatial properties of the source.  We compare different POVMs, finding the optimal separable POVM, as well as a POVM which is close to optimal for a large range of parameters.  This near optimal POVM has the advantage that its implementation does not require knowledge of the parameters, and therefore it can be implemented {\change at the outset} and without the need to adaptively change the POVM.   

In order to compare various schemes we use as a quantifier the cost function
\begin{align}\label{eq:cost}
V_{ M}({\bf I}_Q) = \tr{ {\bf I}_Q {\bf I}_C^{-1}(M) } \,,
\end{align}
where ${\bf I}_Q$ is the quantum Fisher information and ${\bf I}_C^{-1} $ is the inverse of the classical Fisher information matrix for a given POVM, $M$.  In general any positive matrix can be used as the weight matrix in the cost function.  The reason for this particular choice of weight matrix is that it maximises the average fidelity between the estimated state and the actual state \cite{Bagan2006} and is therefore the natural choice for state estimation.  We would expect that a measurement which is simultaneously optimal for all parameters should achieve $V({\bf I}_Q) = d${\change , where $d$ is the number of parameters}.  However, when the SLDs do not commute and we restrict ourselves to separable measurements, we find $V({\bf I}_Q) > d$.

{\change \subsection{Two Spatial Mode State}}

{\change \noindent The state $\rho$ now describes the composite system of two spatially separated volumes, see Fig.~\ref{fig:state2}.}  The addition of a secondary spatial mode changes the covariance matrix in the following way:
\begin{align}\label{eq:CM_2Spat_mo}
\Sigma =
\overset{M}{\underset{ i=1}{\oplus} }\begin{pmatrix}
0 &  c_1^{(i)} & 0 & b_i \\
c_1^{(i)} & 0 & b_i^* & 0 \\
0 & b_i^* & 0 & c_2^{(i)}  \\
b_i & 0 & c_2^{(i)} & 0
\end{pmatrix}  ,
\end{align}
where $\smash{b_i = \braket{\hat{a}_2^{(i) \dagger} \hat{a}_1^{(i)}} }$, $\smash{c_j^{(i)} =\braket{\hat{a}_j^{(i) \dagger} \hat{a}_j^{(i)}} + 1/2} $, and $\smash{\hat{a}^{(i)}_j} $ is the annihilation operator in the spatial mode $j$ and the spectral mode $i$.  To determine the elements $b_i$ and $b_i^*$ we first note that the complex degree of coherence between the two spatial modes is defined as
\begin{align}
\gamma_{12}^{(i)} = \frac{\braket{\hat{a}_2^{(i) \dagger} \hat{a}_1^{(i)}} }{\left[\braket{n^{(i)}_{1}} \braket{n^{(i)}_{2}} \right]^{\frac{1}{2}}} \, ,
\end{align}
which allows us to write $\smash{b_i = [ \braket{n^{(i)}_{1}} \braket{n^{(i)}_{2}}]^{\frac{1}{2}} \gamma^{(i)}_{21}}$ and $\smash{b_i^* = [\braket{n^{(i)}_{1}} \braket{n^{(i)}_{2}}]^{\frac{1}{2}} \gamma^{(i)}_{12}}$. 
We now make use of the van Cittert-Zernike theorem, which states that the far-field complex degree of coherence of a spatially incoherent source is proportional to the Fourier transform of the intensity distribution \cite{MandelandWolf1995}.  Therefore the correlations between the two modes provide information about the Fourier transform of the source distribution.  
 
In Section \ref{sec:Temp_est} we saw that the variables $\smash{\braket{n^{(i)}_{j}}}$ can convey only information about the temperature of the source and $\kappa$.  In contrast, the variables $\gamma^{(i)}_{12}$ convey detailed information about the distribution of the source.  Since the Fourier transform is an injective mapping, knowledge about the Fourier transform over the entire far-field plane can be used to exactly reconstruct the source intensity distribution.  Restricting the detection area of the state will result in incomplete information about the Fourier transform and therefore limits the resolution of a reconstructed intensity distribution.  In this paper we restrict our attention to the optimal estimation of the state consisting of two spatial modes for simplicity. 

We make an additional simplification and assume that the average photon number in both spatial modes is approximately equal, $\smash{\braket{n^{(i)}_{1}} = \braket{n^{(i)}_{2}}}$.  We expect this assumption to hold in the far-field of an isotropic emitter.  For brevity we will also drop the superscript $(i)$ since each frequency mode is independent and therefore can be optimised separately. 

{\change \subsection{Optimal Estimators of Spatial Parameters}}

\noindent Under this assumption we calculate the SLDs for the parameters $\btheta = (\braket{n}, |\gamma | , \phi )^{\rm T}$, where $\smash{\gamma_{12} = |\gamma |  \exp(i \phi)}$, which are found to be of the form
\begin{align}\label{eq:SLD}
\mathcal{L}_j & = P_j \hat{n}_{\rm tot} + Q_j \hat{a}^{\dagger}_1 \hat{a}_2 + Q^{*}_j \hat{a}^{ \dagger}_2 \hat{a}_1 + R_j \id ,
\end{align}
where $\smash{\hat{n}_{\rm tot} = \hat{n}_1 + \hat{n}_2}$ and explicit expressions for $\smash{P_j}$, $\smash{Q_j} $, and $\smash{R_j}$ are given in App.~\ref{app:SLDs}.  As we show in App.~\ref{app:SLDs}, the commutators for these operators are $[\mathcal{L}_1,\mathcal{L}_2]=0$, $[\mathcal{L}_{1},\mathcal{L}_3] \neq 0$, $[\mathcal{L}_{2},\mathcal{L}_3] \neq 0$ and therefore we cannot find a simultaneous eigenbasis for all three operators.  This rules out the possibility of simultaneously measuring in the eigenbases of the SLDs to achieve a simultaneously optimal estimate of each parameter.  Since they do however satisfy the condition Eq.~\eqref{eq:expcomm} (see App.~\ref{app:SLDs}), it is possible that a collective measurement exists which attains the QCRB asymptotically \cite{Gill2011}.  

In this paper we will not consider collective measurements due to the immense technical obstacles.  Instead we consider only the class of measurements which are separable.  To find the optimal POVM, we first determine the operators $\smash{\bX = {\bf I}_Q^{-1} \bL}$, where we have defined $\smash{\bX = (X_1,X_2,X_3)^{\rm T}}$ and $\smash{\bL = (\mathcal{L}_1,\mathcal{L}_2,\mathcal{L}_3)^{\rm T}}$.  {\change This is the multi-parameter extension of Eq.~\eqref{eq:Xop} and has the desirable property that ${\rm Var}(X_i) = [{\bf I}_Q^{-1}]_{ii}$.}  

Determining the commutators of these operators, we find $[X_1,X_2]=[X_1,X_3]=0$, $[X_2,X_3]\neq0$ (see App.~\ref{app:SLDs}) and therefore $X_2$, $X_3$ cannot be simultaneously measured.  Next we calculate the quantum Fisher information from the covariance matrix Eq.~\eqref{eq:CM_2Spat_mo} and determine the classical Fisher information for measurements of the operators $X_2$, $X_3$.  We find,
\begin{align}
{\bf I}_Q & =
\begin{pmatrix}
[{\rm I}_Q]_{11} & [{\rm I}_Q]_{12} & 0 \\
[{\rm I}_Q]_{12} & [{\rm I}_Q]_{22} & 0 \\
0 & 0 & [{\rm I}_Q]_{33} \\
\end{pmatrix} \, , \label{eq:IQ} \\
{\bf I}_C(X_2) & =
\begin{pmatrix}
[{\rm I}_Q]_{11} & [{\rm I}_Q]_{12} & 0 \\
[{\rm I}_Q]_{12} & [{\rm I}_Q]_{22} & 0 \\
0 & 0 & \makebox[\widthof{$[{\rm I}_Q]_{33}$}][c]{0} \\
\end{pmatrix} \, , \label{eq:IC2} \\
{\bf I}_C(X_3) & =
\begin{pmatrix}
[{\rm I}_Q]_{11} & \delta_1 & 0 \\
\delta_1 & \makebox[\widthof{$[{\rm I}_Q]_{22}$}][c]{$\delta_2$} & 0  \\
0 & 0 &  [{\rm I}_Q]_{33} \\
\end{pmatrix} \, , \label{eq:IC3}
\end{align}
where $\delta_1 \ll  [{\rm I}_Q]_{12} $ and $\delta_2 \ll  [{\rm I}_Q]_{22} $.

We notice that a slight asymmetry exists between the operators $X_2$ and $X_3$.  Whilst measurements of $X_2$ do not provide information about $\phi$ (since $[{\bf I}_C(X_2)]_{33} = 0$), measurements of $X_3$ do provide a small amount of information about $|\gamma |$ ($\delta_2, \delta_3 \neq 0$).  In App.~\ref{app:OWMS} we argue that the optimal measurement scheme is given by the solution to the following minimization:
\begin{align}\label{eq:VarO}
\min_p \tr{{\bf I}_Q (p \, {\bf I}_C(X_2) + (1-p) {\bf I}_C(X_3))^{-1}}.
\end{align}
This scheme can be interpreted as probabilistically choosing to measure $X_2$ or $X_3$ with probabilities $p$ and $1-p$ respectively.  The minimisation in Eq.~\eqref{eq:VarO} ensures that we choose the optimal weighting with respect to $X_2$ and $X_3$.  We refer to this scheme as the weighted measurement scheme.  

\begin{figure}[t]
\centering
\begin{tikzpicture}[]
  \pgftext{\includegraphics[width=0.8\columnwidth]{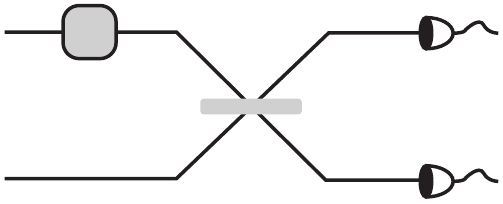}} at (0,0);
  \node at (1,-0.04) {BS};
  \node at (-2.09,0.9) {\large $\varphi$};
\end{tikzpicture}
\caption{Measurement scheme with variable phase shift $\varphi$.  Choosing different values for the phase shift allows optimal estimation of the parameters $\braket{n}$, $|\gamma |$, and $\phi$.}\label{fig:MS}
\end{figure}

Due to the simple quadratic forms of the SLDs, and therefore also the operators $\bX$, we can determine the eigenmodes of $X_2$ and $X_3$.  We find that both operators can be measured using the scheme shown in Fig.~\ref{fig:MS}, where the value of $\varphi$ is different for $X_2$ and $X_3$.  To measure $X_2$ requires the choice $\varphi = \phi$ and $X_3$ can be achieved by setting $\varphi = \phi - \pi/2$.  An optimal measurement therefore requires switching between the two distinct phase settings.  As discussed above, the probability with which each measurement is made should also be optimised.  The exact value of the probability $p$ will depend on the exact values of the parameters.  Since this scheme clearly requires knowledge of the parameters we are attempting to measure, it can only be implemented in an adaptive scenario.  However, since we now know the optimal measurement, we can use this to compare other, simpler schemes, with the aim to find a scheme that is close to optimal but is independent of the parameters.
\begin{figure}[h!]
\begin{tikzpicture}[]
  \node at(-0.536,-6.5) {\includegraphics[width=0.8\columnwidth]{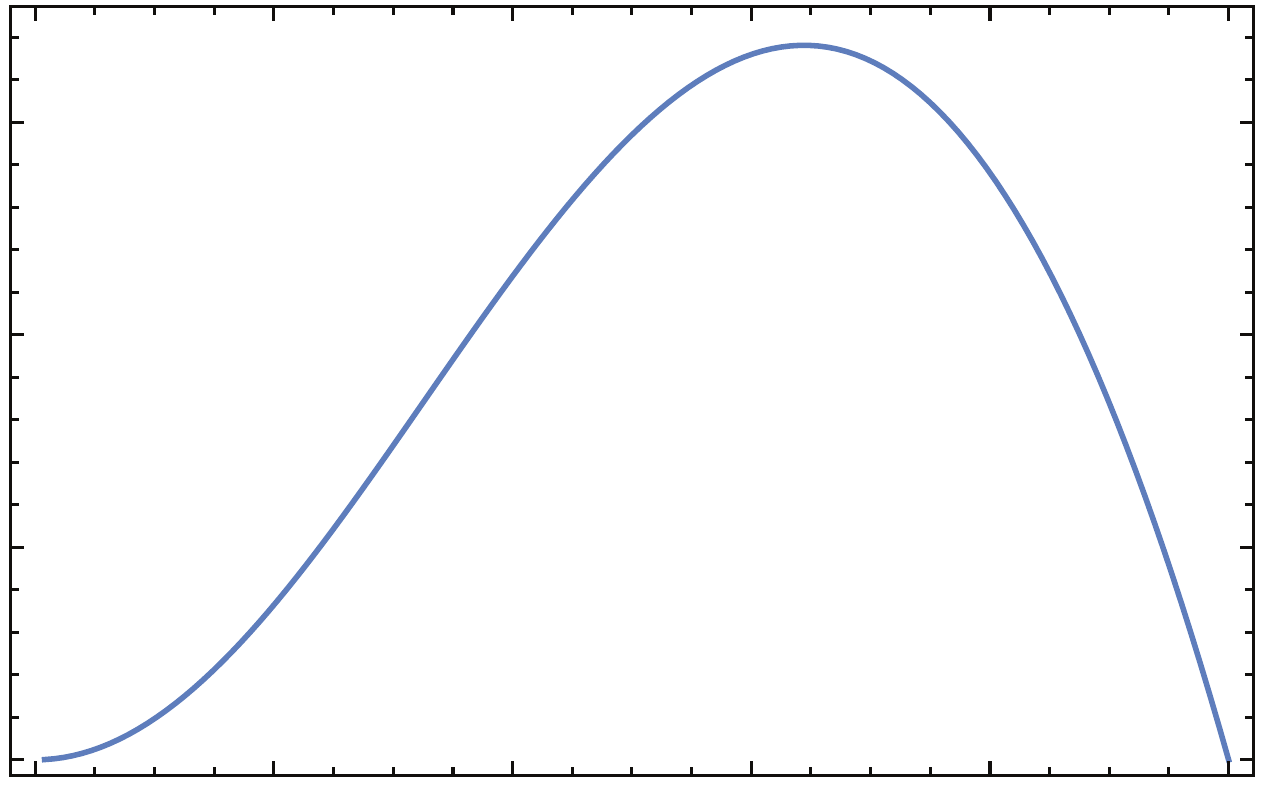}};	
  \node at(-0.78,0) {\includegraphics[width=0.7\columnwidth]{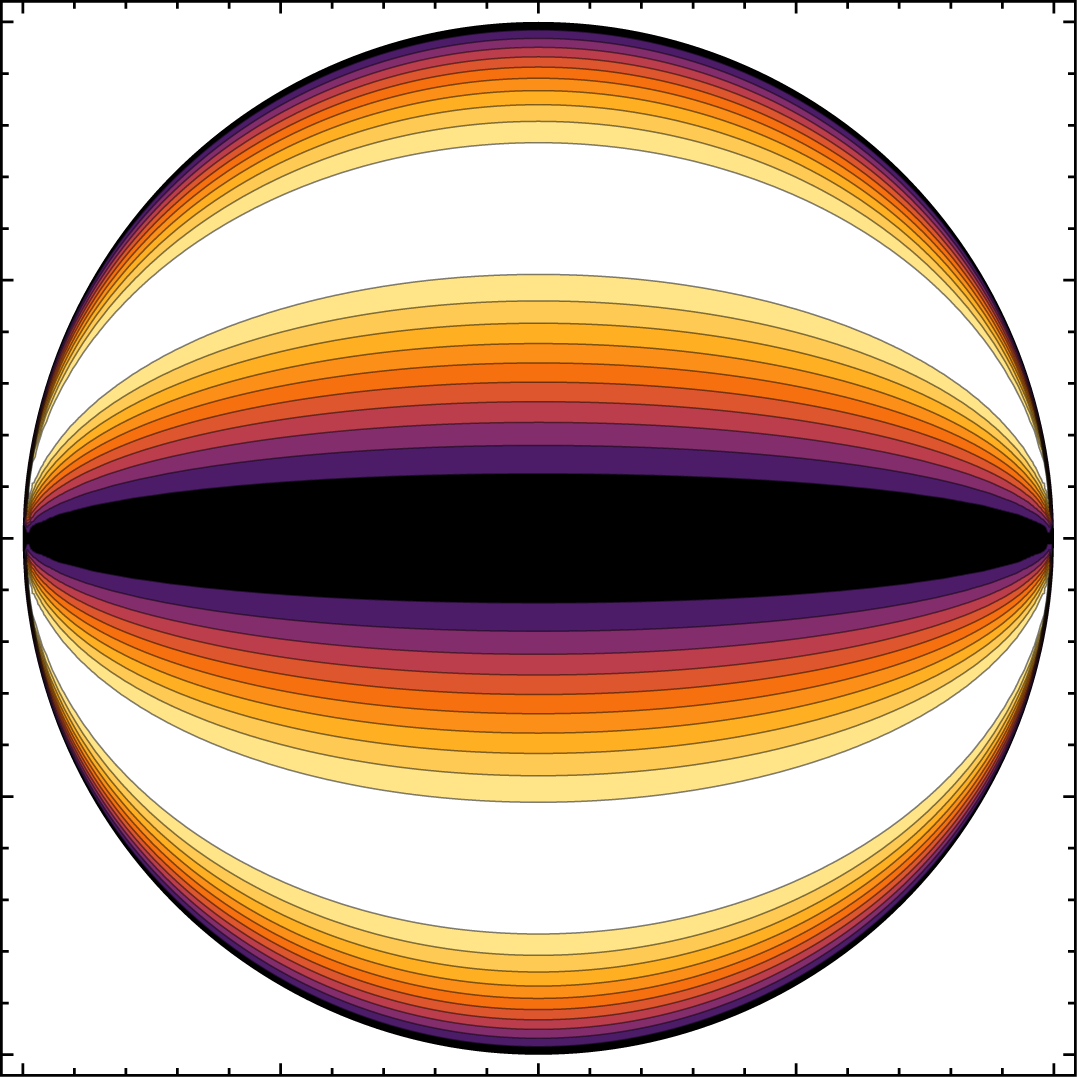}};
  \node at(2.5,0) {\includegraphics[width=0.04\columnwidth]{bar1.png}};
  \draw[dashed] (-0.78,0)--(-0.78,2.8);
  \node at (-4.75,3.1) {a)};
  \node at (-0.78,-3.1) {0};
  \node at (0.6,-3.1) {0.5};
  \node at (1.96,-3.1) {1};
  \node at (-2.18,-3.1) {-0.5};
  \node at (-3.5,-3.1) {-1};
  \node at (-0.83 ,-3.7) {$|\gamma | \, \cos \phi$};
  
  \node[anchor=east] at (-3.65,0) {0};
  \node[anchor=east] at (-3.65,1.36) {0.5};
  \node[anchor=east] at (-3.65,2.75) {1};
  \node[anchor=east] at (-3.65,-1.36) {-0.5};
  \node[anchor=east] at (-3.65,-2.75) {-1};
  \node [rotate=90] at (-4.65,0){$|\gamma | \, \sin \phi$};
  \node[anchor=west] at (2.65,-1.44) {3};
  \node[anchor=west] at (2.65,-0.62) {6};
  \node[anchor=west] at (2.65,0.23) {9};
  \node[anchor=west] at (2.65,1.03) {12};
  \node[anchor=west] at (2.65,1.85) {15};
  \node[anchor=west] at (2.1,-2.5) {$\times 10^{-3}$};
  
  \node at (-4.75,-4.3) {b)};
  \node at (-3.61,-8.75) {0};
  \node at (-2.37,-8.75) {0.2};
  \node at (-1.14,-8.75) {0.4};
  \node at (0.11,-8.75) {0.6};
  \node at (1.31,-8.75) {0.8};
  \node at (2.55,-8.75) {1};
  \node at (-0.55,-9.4) {$|\gamma |$};
  
  \node[anchor=east] at (-3.8,-7.28) {5};
  \node[anchor=east] at (-3.8,-6.19) {10};
  \node[anchor=east] at (-3.8,-5.11) {15};
  \node[rotate=90] at (-4.75,-6.5) {$V_{\rm op}({\bf I}_Q) / V_{\rm FT}({\bf I}_Q) \,\, (\times 10^{-3}$)};
  
\end{tikzpicture}
\caption{$V_{\rm op}({\bf I}_Q) / V_{\rm FT}({\bf I}_Q)$ for the measurement scheme involving only a Fourier transform and photon counting.  a) We see that this scheme is suboptimal for the entire parameter range $0\leq |\gamma | \leq 1$, $0 \leq \phi < 2\pi$ and in fact never achieves more than 1.7 per cent of the performance of the optimal estimator.  b) Shows a cut through fig a) along the dashed line.  In this plot we set $\braket{n}=0.01$. }\label{fig:FT}
\end{figure}

{\change \subsection{Fixed Phase Scheme}}

We first consider how well the parameters can be estimated using a Fourier transform and photon counting on the two input modes.  This is of interest for optics because it closely approximates the action of lensing and intensity measurements, which is by far the most commonly used method in imaging.  The photon count probability distributions for both the state $\rho$ and the state after the action of the Fourier transform are derived in App. \ref{app:PCPD}.  

We find that without the Fourier transform the photon count distribution is independent of the phase $\phi$ and therefore cannot be used to estimate $\phi$.  However, once the Fourier transform has been applied to the input modes all three parameters can be estimated.  In Fig.~\ref{fig:FT} we plot the ratio of the cost function for the Fourier transform scheme, $V_{\rm FT}({\bf I}_Q)$, to the optimal scheme for the entire range of the parameters $0\leq |\gamma | \leq 1$, and $0 \leq \phi < 2\pi$, which follows from the definition of $\gamma_{12}$ and the restriction $0 \leq |\gamma_{12}| \leq 1$.  We see that this scheme performs far below the optimal scheme and is therefore not an efficient measurement for determining the spatial configuration of the source.

{\change \subsection{Random Phase Scheme}}

\begin{figure}[h!]
\begin{tikzpicture}[]
  \node at(-0.536,-6.5) {\includegraphics[width=0.8\columnwidth]{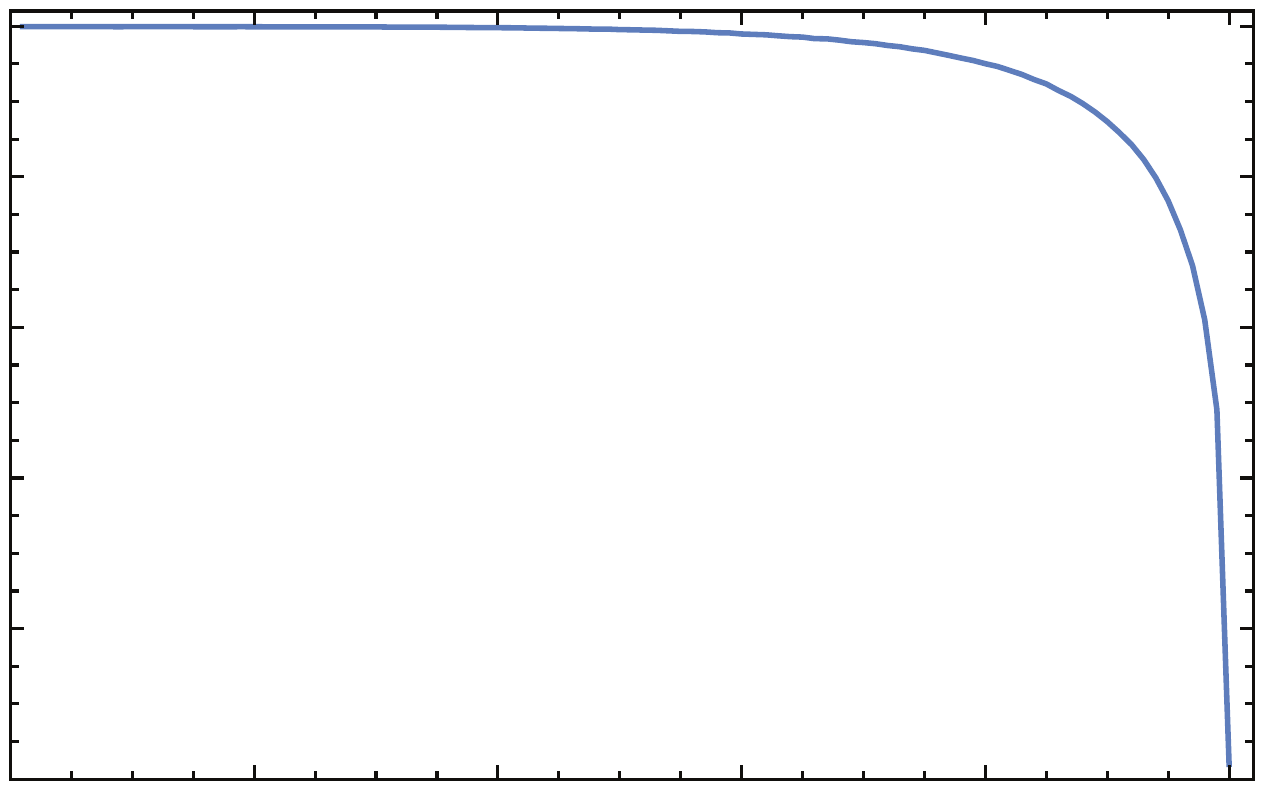}};	
  \node at(-0.78,-0.1) {\includegraphics[width=0.7\columnwidth]{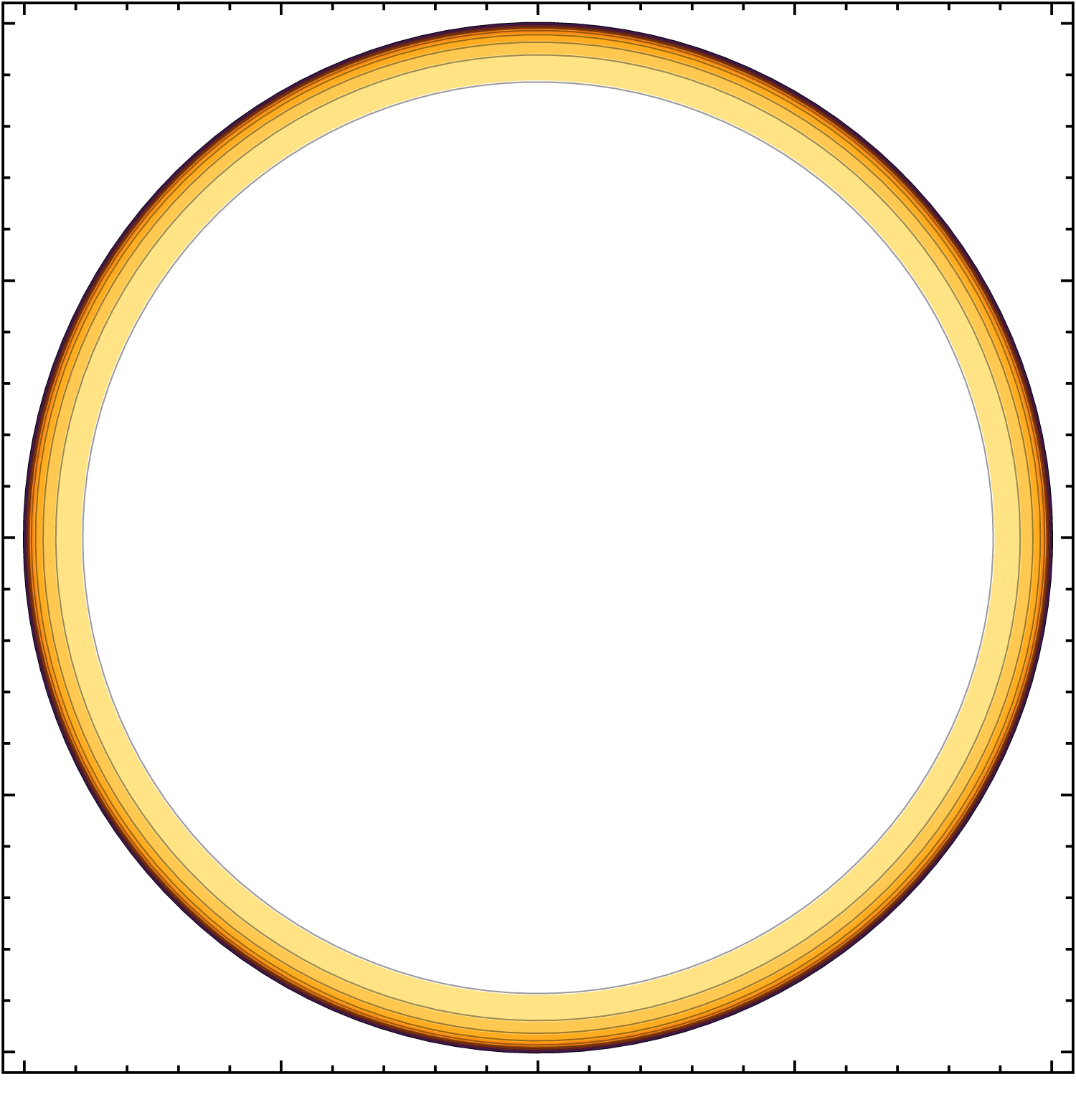}};
  \node at(2.5,0) {\includegraphics[width=0.04\columnwidth]{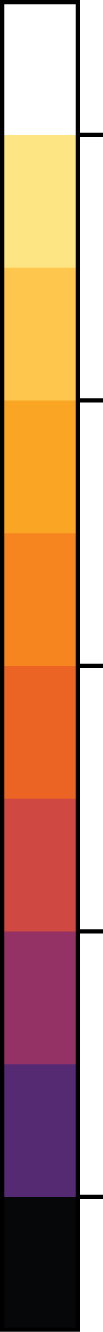}};
  \node at (-4.75,3.1) {a)};
  \node at (-0.78,-3.1) {0};
  \node at (0.6,-3.1) {0.5};
  \node at (1.96,-3.1) {1};
  \node at (-2.18,-3.1) {-0.5};
  \node at (-3.5,-3.1) {-1};
  \node at (-0.83 ,-3.7) {$|\gamma | \, \cos \phi$};
  
  \node[anchor=east] at (-3.65,0) {0};
  \node[anchor=east] at (-3.65,1.36) {0.5};
  \node[anchor=east] at (-3.65,2.75) {1};
  \node[anchor=east] at (-3.65,-1.36) {-0.5};
  \node[anchor=east] at (-3.65,-2.75) {-1};
  \node [rotate=90] at (-4.65,0){$|\gamma | \, \sin \phi$};
  \node[anchor=west] at (2.65,-1.67) {0.1};
  \node[anchor=west] at (2.65,-0.835) {0.3};
  \node[anchor=west] at (2.65,0) {0.5};
  \node[anchor=west] at (2.65,0.835) {0.7};
  \node[anchor=west] at (2.65,1.67) {0.9};
  
  \node at (-4.75,-4.3) {b)};
  \node at (-3.74,-8.75) {0};
  \node at (-2.46,-8.75) {0.2};
  \node at (-1.2,-8.75) {0.4};
  \node at (0.04,-8.75) {0.6};
  \node at (1.3,-8.75) {0.8};
  \node at (2.55,-8.75) {1};
  \node at (-0.55,-9.4) {$|\gamma |$};
  
  \node[anchor=east] at (-3.8,-8.45) {0};
  \node[anchor=east] at (-3.8,-7.7) {0.2};
  \node[anchor=east] at (-3.8,-6.95) {0.4};
  \node[anchor=east] at (-3.8,-6.15) {0.6};
  \node[anchor=east] at (-3.8,-5.38) {0.8};
  \node[anchor=east] at (-3.8,-4.59) {1};
  \node[rotate=90] at (-4.75,-6.5) {$V_{\rm op}({\bf I}_Q) / V_{\rm RP}({\bf I}_Q)$};
  
\end{tikzpicture}
\caption{The ratio of $V_{\rm op}({\bf I}_Q)$ to $V_{\rm RP}({\bf I}_Q) $.   The random phase scheme is seen to be close to optimal unless $|\gamma |$ is close to one.  For this figure we randomly sampled 1000 phases $\varphi$ and determined the inverse of the classical Fisher information for such a set of measurements.  Since the exact performance is affected by the exact values of $\varphi$ we averaged each point for over 400 trials to remove most of the statistical noise.  The inset shows the performance over the whole range, which shows that the performance of the random phase scheme is independent of $\phi$.}\label{fig:RP}
\end{figure}
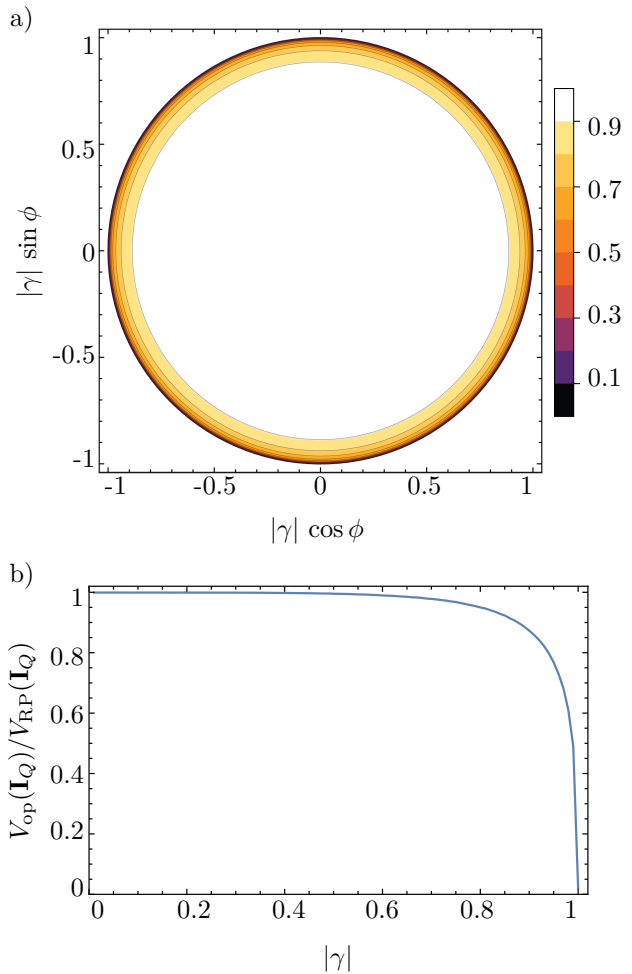

The second scheme we consider is a scheme where we randomly select a phase shift, uniformly over the range $\varphi \in [0,2 \pi)$, then pass the modes through a beamsplitter and measure in the number basis.  We call this scheme the random phase (RP) scheme.  In Fig.~\ref{fig:RP} we plot the ratio $V_{\rm op}({\bf I}_Q) / V_{\rm RP}({\bf I}_Q)$ to compare the random phase scheme to the optimal measurement scheme.
We find that the random phase scheme is very close to optimal for most of the parameter range $0\leq |\gamma | \leq 1$, $0 \leq \phi < 2\pi$.  The advantage of the random phase scheme is that, in implementing this scheme, we do not require the values of the parameters.  This therefore means that this POVM can be implemented before we have acquired any information about the parameters.  Surprisingly, even without this information, the random phase scheme performs very efficiently.   

As discussed above, the truly optimal POVM requires us to choose measurements of $X_2$ with probability $p$ and $X_3$ with probability $(1-p)$.  Although this achieves the optimal performance it requires knowledge of the parameter values and therefore can only be implemented in an adaptive way.  We expect that an approximately optimal scheme can be achieved by first making measurements using the random phase scheme, and when the estimates of the parameters have reached a sufficient precision the optimal scheme can be used to further enhance the measurement precision.

\section{Discussion and Conclusions}

\noindent In this paper we have derived a complete description of states originating in far-field blackbody sources {\change and observed over a finite detection area in the far-field.}  {\change This description is complete in the sense that it encompasses all parameters that determine the state, including} nuisance parameters that must be estimated in order to estimate the temperature of the blackbody {\change and its spatial configuration}.  We found that it is necessary to attempt to estimate these nuisance parameters in order to estimate the temperature and to do so requires photon counting to be performed on at least two spectral modes.

In order to derive our results, it was necessary to make certain assumptions about the arrangement of the source and its relation to the state $\rho$.  In the paper we assumed that the transverse area of the source is much less than one coherence area.  We demonstrated that this gave a bound on the frequencies we can measure, $10^{-14} \nu \ll \cot \vartheta$, where $\vartheta$ is the angular size of the source.  We also found a linear relationship between the optimal pair of frequencies and the temperature.  Putting these results together allows us to find the following bound {\change $\cot \vartheta \gg \epsilon \, T$, where $\epsilon \approx 10^{-3} {\rm K}^{-1}$ }which shows that the optimal frequencies for very hot sources ($\sim10^5 - 10^6$K) should fall in the range of acceptable frequencies for far-field objects.

Interestingly, we found that the estimation of general spatial parameters requires that the states are observed over at least two spatially separated modes.  The introduction of a second spatial mode introduces a complex parameter which determines the coherence between the spatial modes and can be identified as the Fourier transform of the intensity distribution of the source.  We then identified the optimal separable measurement for the resulting three parameter estimation problem.  

The optimal strategy depends upon the exact value of the parameters we wish to estimate, and therefore can only be implemented in an adaptive setting.  However, we also identified another strategy, the random phase scheme, which performs close to optimal for a wide range of the parameters (see Fig.~\ref{fig:RP}).  The implementation of the random phase scheme is independent of the parameter values and therefore has the advantage of not requiring an adaptive arrangement.  We can then suppose that an asymptotically optimal scheme can be achieved by starting with the random phase scheme, using this to find a reasonable estimate of the parameters, and then implementing the optimal scheme to further increase the precision of the estimates.

We also considered the performance of the measurement scheme consisting of simply a Fourier transform and photon counting.  This is actually a special case of the measurement scheme in Fig.~\ref{fig:MS}, with $\varphi$ fixed at a constant value (namely zero).  Our motivation is that this scheme most closely resembles a typical imaging scheme, with the role of the Fourier transform being provided by a lens and photon counting provided by a CCD or other photosensitive surface.  We find that this scheme performs poorly over the entire parameter range.  

Our results expose simple imaging schemes as far from optimal, even in the simplistic two mode settings.  We have shown how spatial information is conveyed to the far-field, and how we can optimally extract this information for the simplest case of two modes.  We expect that this work will enable further research into quantum optimal imaging and helps to answer the two open questions: 1) What are the optimal measurements for measuring spatial features of radiating sources?  2) How well do standard imaging techniques compare to optimal schemes? 

\section*{Acknowledgements}

\noindent The authors would like to thank Mark Howard, Madalin Guta, and Mankei Tsang for helpful discussions relating to the issues addressed in this paper.  MEP and PK acknowledge EPSRC for funding via the Quantum Communications Hub.  ETC is supported by the EPSRC (EP/M024261/1).

\begin{appendix}

\section{Symmetric Logarithmic Derivatives of $\braket{n}$, $|\gamma |$, and $\phi$}\label{app:SLDs}

\noindent Here we give the explicit forms of the variables $P_j$, $Q_j$, and $R_j$, defined in Eq.~\eqref{eq:SLD}, which determine the SLDs for the three parameters $\btheta = (\braket{n}, |\gamma |, \phi )^{\rm T}$.  They are:
\begin{align}
P_{\braket{n}} & = \frac{\braket{n}+1}{\braket{n}\left[ \braket{n}^2 |\gamma |^2 - (\braket{n} -1)^2 \right]} \label{eq:Xn} \\
Q_{\braket{n}} & = \frac{|\gamma | {\rm e}^{-i \phi}}{\braket{n}\left[ \braket{n}^2 |\gamma |^2 - (\braket{n} -1)^2 \right]} \\
R_{\braket{n}} & = \frac{2 \braket{n} (|\gamma |^2 \braket{n} -\braket{n} -1)}{\braket{n}\left[ \braket{n}^2 |\gamma |^2 - (\braket{n} -1)^2 \right]}  \\
P_{|\gamma |} & = \frac{2 \braket{n}+1}{(|\gamma |^2 - 1)\left[ \braket{n}^2 |\gamma |^2 - (\braket{n} -1)^2 \right]}  \\
Q_{|\gamma |} & = \frac{{\rm e}^{-i \phi} (1+ \braket{n} + |\gamma |^2 \braket{n}^2)}{(|\gamma |^2 - 1)\left[ \braket{n}^2 |\gamma |^2 - (\braket{n} -1)^2 \right]} \\
R_{|\gamma |} & = \frac{2|\gamma | \left[ \braket{n}^2 (|\gamma |^2 - 1) \right]}{(A^2 - 1)\left[ \braket{n}^2 |\gamma |^2 - (\braket{n} -1)^2 \right]} \\
P_{\phi} & = 0  \\
Q_{\phi} & = i |\gamma | {\rm e}^{-i \phi} \\
R_{\phi} & = 0 \label{eq:Zphi} \, .
\end{align}
Taking the commutator of each of the SLDs pairwise, we find $[\mathcal{L}_{\braket{n}}, \mathcal{L}_{|\gamma |}] = 0$, and $[\mathcal{L}_{\braket{n}}, \mathcal{L}_{\phi}] , [\mathcal{L}_{|\gamma |}, \mathcal{L}_{\phi}] \propto \hat{n}_1 - \hat{n}_2$.  As we show in the main paper, the observables that provide optimal information about each parameter independently are the set $\bX$, given by  
\begin{align}
X_i = \left[{\bf I}_Q^{-1} \boldsymbol{\mathcal{L}} \right]_i \, .
\end{align}
From Eqs.~\eqref{eq:Xn} to \eqref{eq:Zphi} we can determine the commutators of the $\bX$ operators.  These are $[X_{\braket{n}}, X_{|\gamma |}] = [X_{\braket{n}}, X_{\phi}] = 0$, and $[X_{|\gamma |}, X_{\phi}] \propto \hat{n}_1 -\hat{n}_2 $.  Therefore we cannot measure $X_{|\gamma |}$ and $X_{\phi}$ simultaneously.

\section{Optimality of Weighted Measurement Scheme}\label{app:OWMS}

\noindent Here we present evidence of the optimality of the weighted measurement scheme presented in Sec.~\ref{sec:TSM}.  We appeal to the results of \cite{Gill2000} where the authors show
\begin{align}\label{eq:VLB1}
\tr{{\bf I}_Q^{-1} {\bf I}_C} & \leq D-1 \,  ,
\end{align}
where $D$ is the dimension of the Hilbert space.  For a $d$ parameter problem, it immediately follows that
\begin{align}
\label{eq:VLB2}
\tr{{\bf I}_Q {\bf I}_C^{-1} } & \geq \frac{d^2}{D-1} \, .
\end{align}
In our setting, the dimension is infinite.  However, since the state consists of thermal modes, which typically have very small average photon numbers \cite{MandelandWolf1995}, we can truncate the state to the 0, 1 photon basis with low truncation error.  Using this $D=3$ approximation {\change (the basis of $\ket{0,0}$, $\ket{0,1}$, $\ket{1,0}$)} suggests that 4.5 is a good lower bound on $\tr{{\bf I}_Q {\bf I}_C^{-1}}$.  However, there is no reason to believe it is attainable. In the main text, we reported that we can achieve just below 5.  Our approach is to choose between the optimal measurements with some probability.  Below we show that for qubits, such an approach is optimal and saturates the lower bound of Eq.~\eqref{eq:VLB2}.  However, for qutrits and low photon number Gaussian states, our numerical investigation indicate that it is not possible to saturate Eq.~\eqref{eq:VLB2}, though we come very close.  

The quantity $\tr{{\bf I}_Q {\bf I}_C^{-1} }$ is invariant under reparameterisation and we can always work in parameterisation where ${\bf I}_Q$ is diagonal~\cite{Gill2000}. If we make the optimal measurement for parameter $i$, we obtain a classical Fisher information ${\bf I}_C^{(i)}$ such that $\smash{[{\bf I}_C^{(i)}]_{i,i}=[{\bf I}_Q]_{i,i}}$.  We may obtain information about other parameters and so
\begin{equation}
	{\bf I}_C^{(i)} \geq \tilde{{\bf I}}_C^{(i)},
\end{equation}
where $\tilde{{\bf I}}_C^{(i)} = {\rm diag} (0 
, \dots, [{\bf I}_Q]_{ii}, \dots ,0)$.  For a $d$ parameter problem, we propose a scheme were each optimal observable $X_i$ is measured on a fraction $1/d$ of the samples.  Then we obtain 
\begin{equation}
	{\bf I}_C = \sum_j \frac{1}{d} {\bf I}_C^{(j)} \geq  \sum \frac{1}{d} \tilde{{\bf I}}_C^{(i)}  = \frac{1}{d} {\bf I}_Q .
\end{equation}
Therefore, 
\begin{equation}
 \tr{{\bf I}_Q {\bf I}_C^{-1} } \leq d \,\tr{ \id }  = d^2 .
\end{equation}
For a qubit, Eq.~\eqref{eq:VLB2} gives a lower bound of $d^2$ and so we see such a scheme is optimal. For a general qutrit problem, we have
\begin{equation}
  \frac{d^2}{2}  \leq \tr{{\bf I}_Q {\bf I}_C^{-1} } \leq  d^2 
\end{equation}
with the actual optimal depending on the exact nature of the problem. For a three parameter problem, we have a lower bound of 4.5, and the above scheme achieves 9 or better.   In the problem of interest to us, some of our optimal measurements commute (see App.~\ref{app:SLDs}) and so we expect to do much better than 9.

Inspired by the above, to find the exact optimum for our problem we allow for minimisation with respect to the relative probability that we measure $X_A$ and $X_{\phi}$.  Denoting the probability that we measure $X_A$ as $p$ and using the approximation $\delta_1=\delta_2 = 0$, we find   
\begin{align}
\min_{p} \,\,  \tr{{\bf I}_Q (p {\bf I}^{(2)}_C + (1-p) {\bf I}^{(3)}_C)^{-1}} & = \nonumber \\  
\min_{p} \,\, 1 + \frac{1}{p} + \frac{1}{1-p} & = 5 \, ,
\end{align}
where the value of $p$ that minimises this expression is $p =1/2$.  When $\delta_1, \delta_2 \neq 0$, the value of $p$ that minimises is slightly less than one half and the cost is just less than 5. 

To further support our claim that this scheme is indeed optimal, we perform a numerical optimisation over a subset of POVMs.  In order to reduce the number of parameters to optimise, we search over the truncated Hilbert space of $\ket{0,0}, \ket{0,1},\ket{1,0}$.  We search over the set of POVMs with six elements corresponding to two sets of three orthogonal components.  To perform the optimisation, we apply two unique three dimensional unitaries to the POVM with elements $\{ \ket{0,0}\bra{0,0}, \ket{0,1}\bra{0,1}, \ket{1,0}\bra{1,0}\}$, giving two POVMs with three elements each, $M_1$ and $M_2$.  A six element POVM is constructed by taking $M= p M_1 + (1-p) M_2$, for which we calculate the classical Fisher information.  We then search for the minimum with respect to the cost function $\tr{{\bf I}_Q {\bf I}_C^{-1}}$, where ${\bf I}_C$ is the classical Fisher information associated with measurements of this POVM.  Performing the optimisation across the range $0 \leq |\gamma | \leq 1$, the results are in good agreement with the values obtained for the weighted measurement scheme.   

\section{Photon Count Probability Distributions}\label{app:PCPD}

\noindent To determine the photon count probability distributions we first notice that the state with covariance matrix Eq.~\eqref{eq:CM_2Spat_mo} can be obtained by the action of the unitary 
\begin{align}
U(\phi) = \frac{1}{\sqrt{2}} \begin{pmatrix} -{\rm e}^{-i \phi} & {\rm e}^{-i \phi} \\
1 & 1 \end{pmatrix},
\end{align}
acting on the mode space of the two mode, thermal state with covariance matrix 
\begin{align}
\begin{pmatrix} 0 & x_1 +\frac{1}{2} \\
x_1 +\frac{1}{2} & 0 \end{pmatrix} \oplus \begin{pmatrix} 0 & x_2 +\frac{1}{2} \\
x_2 +\frac{1}{2} & 0 \end{pmatrix},
\end{align}
where $x_1 = \braket{n}(1-|\gamma |)$, $x_2 = \braket{n}(1+|\gamma |)$.  Since the modes are independent, the photon count probability distribution is simply given by 
\begin{align}
p_{\rm in}(n_1,n_2) = \frac{x_1^{n_1}}{(1 + x_1)^{n_1+1}} \frac{x_2^{n_2}}{(1 + x_2)^{n_2+1}},
\end{align}
and the state can be expressed in the number basis as
\begin{align}
\rho_{\rm in} = \sum_{n_1,n_2 = 0}^{\infty}p_{\rm in}(n_1,n_2) \ket{n_1,n_2} \bra{n_1,n_2}.
\end{align}
Making use of the relation $\ket{n_i} = n_i^{-{\frac{1}{2}}} (\hat{a}_i^{\dagger})^{n_i} \ket{0}$, and taking the transformation $[\hat{a}_{\rm in}]_i = U_{ij}^{\dagger}(\phi) [\hat{a}_{\rm out}]_j $, we find the photon counting probability distribution is given by 
\begin{align}
p_{\rm out}(m_1,m_2) = \sum_{n_1=0}^{m_1+m_2} \frac{p_{\rm in}(n_1,m_1+m_2-n_1) }{n_1! (m_1+m_2-n_1)!} \frac{m_1! m_2!}{ 2^{m_1+m_2}} \nonumber \\
\times \left| \sum_{j=0}^{m_1} (-1)^j{ n_1 \choose j}  { m_1+m_2-n_1 \choose m_1 - j} \right|^2,
\end{align}
where we have made use of the binomial theorem.  We notice that this distribution is independent of $\phi$ and therefore cannot be used to obtain estimates of $\phi$.

To obtain the distribution after the application of the Fourier transform, we note that the discrete 2D-Fourier transform is given by $U(0)$, we therefore use the transformation $[\hat{a}_{\rm in}]_i = [(U(0) U(\phi))^{\dagger}]_{ij} [\hat{a}_{\rm out}]_j $ to obtain the distribution
\begin{widetext}
\begin{align}\label{eq:PND_FT}
& p_{\rm out}(m_1,m_2) = \nonumber \\
& \sum_{n_1=0}^{m_1+m_2} \frac{p_{\rm in}(n_1,m_1+m_2-n_1) }{n_1! (m_1+m_2-n_1)!} \frac{m_1! m_2!}{ 4^{m_1+m_2}} \left| \sum_{j=0}^{m_1} (-1)^j{ n_1 \choose j}  { m_1+m_2-n_1 \choose m_1 - j}(1-{\rm e}^{-i \phi})^{m_1+n_1-2j}(1+{\rm e}^{-i \phi})^{m_1-n_1+2j} \right|^2.
\end{align}
\end{widetext}
The presence of $\phi$ in Eq.~\eqref{eq:PND_FT} means that, after the application of the Fourier transform, photon counting can be used to determine $\phi$.  The classical Fisher information for these distributions is simply given by
\begin{align}
{\bf I}_C = \sum_{\substack{m_1,m_2 \\ =0}}^{\infty} \frac{\left(\nabla_{\btheta} p_{\rm out}(m_1,m_2) \right) \left( \nabla_{\btheta} p_{\rm out}(m_1,m_2) \right)^{\rm T}}{p_{\rm out}(m_1,m_2)} ,
\end{align}
where $\nabla_{\btheta} = (\frac{\partial}{\partial \theta_1}, \frac{\partial}{\partial \theta_2}, \dots, \frac{\partial}{\partial \theta_l})^{\rm T}$.

\end{appendix}

\bibliographystyle{apsrev4-1}

\end{document}